\documentclass[conference]{IEEEtran}
\IEEEoverridecommandlockouts
\usepackage{cite}
\usepackage{amsmath,amssymb,amsfonts}
\usepackage{graphicx}
\usepackage{textcomp}
\usepackage{xcolor}
\usepackage{algorithm}
\usepackage{algpseudocode}
\usepackage{svg}

\usepackage{tabularx} 
\usepackage{multirow}

\usepackage{booktabs}   
\usepackage{multirow}   
\usepackage{tabularx}   
\usepackage{amssymb}    
\usepackage{siunitx}

\usepackage{booktabs}
\usepackage{tabularx}
\usepackage{makecell}
\usepackage{siunitx} 
\usepackage{ragged2e} 

\usepackage[caption=false]{subfig}

\def\BibTeX{{\rm B\kern-.05em{\sc i\kern-.025em b}\kern-.08em
    T\kern-.1667em\lower.7ex\hbox{E}\kern-.125emX}}

\usepackage{tikz}

\newcommand\copyrighttext{%
  \footnotesize \textcopyright 2026 IEEE. Personal use of this material is permitted. Permission from IEEE must be obtained for all other uses, in any current or future media, including reprinting/republishing this material for advertising or promotional purposes, creating new collective works, for resale or redistribution to servers or lists, or reuse of any copyrighted component of this work in other works.
}

\newcommand\copyrightnotice{%
\begin{tikzpicture}[remember picture,overlay]
\node[anchor=south,yshift=10pt] at (current page.south)
{\fbox{\parbox{\dimexpr\textwidth-\fboxsep-\fboxrule\relax}{\copyrighttext}}};
\end{tikzpicture}%
}

\newcommand\acceptancenotice{%
\begin{tikzpicture}[remember picture,overlay]
\node[anchor=north,yshift=-20pt] at (current page.north)
{\large \textit{Accepted for publication in the 35th International Conference on Computer Communications and Networks (ICCCN 2026)}};
\end{tikzpicture}%
}

\begin{document}










\title{Privacy-Aware 5G Traffic Classification: A Few-Shot RAN Telemetry Framework}

\author{\IEEEauthorblockN{
Muhammad Enayetur Rahman\IEEEauthorrefmark{1},
Sachin Shetty\IEEEauthorrefmark{1},
Safdar Bouk \IEEEauthorrefmark{1} and
Kimberly Gold\IEEEauthorrefmark{5} 
}
\IEEEauthorblockA{\IEEEauthorrefmark{1} Old Dominion University,
Norfolk, VA. Email: \{mrahm011, sshetty, sbouk\}@odu.edu }

\IEEEauthorblockA{\IEEEauthorrefmark{5} Naval Surface Warfare Center, Crane Division, Crane, IN.
Email: \{kimberly.d.gold2\}.civ@us.navy.mil}
}

 \vspace{1em} 

\maketitle

\acceptancenotice
\copyrightnotice

\begin{abstract}
The proliferation of encrypted traffic and diverse Quality of Service (QoS) demands in 5G networks challenges traditional network management. This paper explores a privacy-preserving approach to traffic classification by leveraging machine learning (ML) models that operate exclusively on Radio Access Network (RAN) metrics. We present a comprehensive study conducted on a fully implemented, end-to-end 5G Standalone (SA) testbed utilizing open-source srsRAN and Open5GS solutions. To guarantee true model generalization across dynamic network states, we evaluate our models under a strict session-based zero-shot protocol. Under this rigorous baseline, classical models like XGBoost achieved an accuracy of 72.87\% but exhibited severe class-level failures when exposed to out-of-distribution network shifts. To bridge this generalization gap, we introduce a few-shot calibration framework that enables rapid domain adaptation using only a 5\% sample of target-domain data. This approach successfully recalibrated a Multi-Layer Perceptron (MLP) architecture, elevating its test accuracy to 86.25\% and fully recovering minority-class detection. Crucially, the calibrated MLP achieves this performance while maintaining an ultra-low inference latency of 0.0006 ms/sample. These findings highlight the model's exceptional efficiency, demonstrating its high suitability for future integration into latency-critical environments, such as real-time MAC-layer scheduling and Open RAN (O-RAN) Near-Real-Time Intelligent Controllers (RIC).
 
\end{abstract}


\section{Introduction}

The surge of diverse, resource-intensive applications in 5G/6G networks creates varied Quality of Service (QoS) demands, making intelligent traffic classification essential for efficient resource management and user QoE \cite{sheraz_6g, computers14010015, miri_demand, foukas_ran_analytics}. However, traditional methods like port-based analysis and Deep Packet Inspection (DPI) are now ineffective due to dynamic ports and the pervasive use of end-to-end encryption \cite{xu2019traffic}. This shift necessitates a privacy-preserving approach, for which Machine Learning (ML) has emerged as a robust alternative. ML models can accurately classify encrypted traffic by learning from the statistical features of communication flows, eliminating the need for sensitive payload inspection \cite{Zhang_micro, bala_open}.

An emerging and highly promising frontier for this approach is to perform classification directly at the Radio Access Network (RAN). The RAN generates a wealth of real-time metrics that inherently reflect an application's network behavior \cite{xavier_icc}. This offers a powerful, privacy-preserving avenue for traffic classification at the network edge. While prior work has validated using ML on network KPIs for high-level tasks like service classification, a detailed investigation linking specific RAN metrics to distinct user applications on a real-world 5G testbed is still needed. Much of the existing research is limited in scope, often confined to a single User Equipment (UE) with a single ML model \cite{khad_lstm} or utilizing simulated data, which may not fully capture real-world network dynamics \cite{xavier_icc, val_qos}. 

Furthermore, applying ML to time-series network data often introduces a pervasive methodological flaw: temporal data leakage caused by randomized train-test splitting. Because RAN telemetry is intrinsically sequential, random sampling allows adjacent, highly correlated time-steps to bleed across evaluation boundaries. Consequently, models frequently report artificially inflated validation accuracies that fail to generalize when deployed in real-world, out-of-distribution scenarios.

To address these gaps, this paper presents a rigorous, privacy-preserving framework for classifying user data based exclusively on RAN metrics, using empirical data from a multi-UE operational 5G testbed. We utilize a comprehensive dataset featuring key RAN parameters such as downlink/uplink Modulation and Coding Scheme (MCS), Signal-to-Noise Ratio (SNR), and Buffer Status Reports (BSR). By explicitly shifting from a flawed randomized-split paradigm to a strict, session-based zero-shot and few-shot evaluation protocol, we accurately quantify the true generalization gap and present a highly efficient calibration method for real-time domain adaptation.

\begin{figure*}[htbp]
    \centerline{\includegraphics[width=0.9\linewidth]{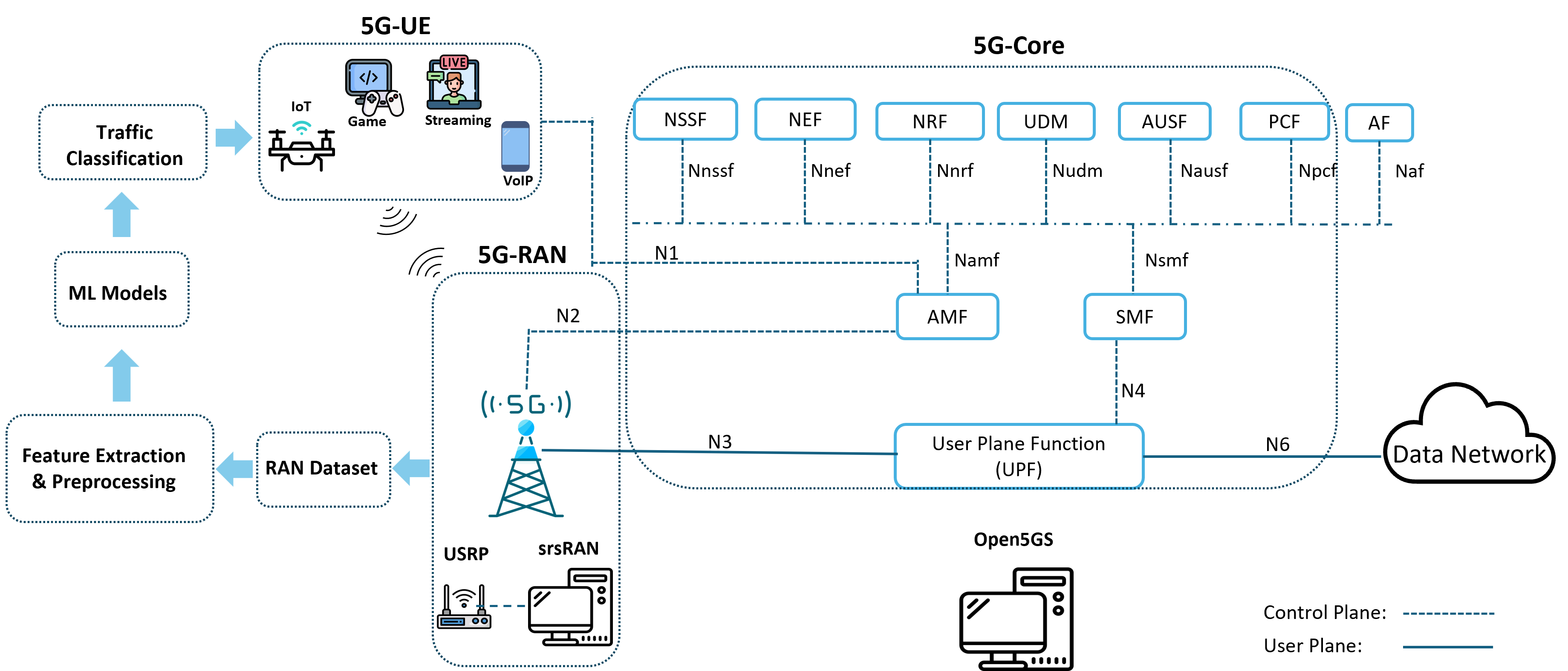}}
    \caption{Overall system model architecture.}
    \label{fig:architecture_label}
\end{figure*}

The primary contributions of this paper are as follows:

\begin{itemize}
    \item We designed and deployed an end-to-end 5G Standalone (SA) testbed utilizing open-source srsRAN and Open5GS to generate a novel dataset. This dataset captures dynamic, time-series RAN telemetry across distinct application profiles, completely bypassing payload inspection.
    \item We identify and analyze critical RAN-level features that are instrumental in accurately categorizing different user traffic types without relying on packet content.
    \item  We implemented a strict session-based zero-shot evaluation protocol to eliminate temporal data leakage and expose the true generalization gap in dynamic networks. We demonstrate that while tree-based models like XGBoost achieve a leading baseline test accuracy of 72.87\%, they suffer from severe class-level failures (such as missing Interactive Gaming traffic entirely) under out-of-distribution shifts.
    \item To resolve this generalization gap, we introduce a few-shot calibration protocol that enables rapid domain adaptation using exactly 5\% of the target-environment samples. We show that this approach successfully recalibrated a Multi-Layer Perceptron (MLP) architecture, elevating its overall accuracy to 86.25\% and fully recovering minority-class detection capabilities.
    \item We present a rigorous analysis of the trade-offs between predictive accuracy, training overhead, and inference latency. We establish that the calibrated MLP classifier - requiring an ultra-low inference latency of just 0.0006 ms/sample - exhibits exceptional efficiency, demonstrating its high suitability for integration into strict, latency-critical environments like Open RAN (O-RAN) Near-Real-Time Intelligent Controllers (RIC).     
\end{itemize}


The paper is organized as follows. Section II details our 5G SA testbed architecture. Section III outlines the experimental methodology. In Section IV, we present and discuss the results of our comparative analysis and feature importance. Finally, Section V concludes the paper, summarizing our findings and outlining future research directions.

\section{System Architecture And Data Collection}

\subsection{System Architecture of a 5G Standalone Network}

Our end-to-end 5G testbed is a software-defined emulation of a 3GPP Release 15 Standalone (SA) network \cite{Ghosh_5g}, built to generate high-fidelity data for ML-driven network analysis. The architecture, shown in Figure 1, is segmented into the User Equipment (UE), Radio Access Network (RAN), and 5G Core (5GC), along with a dedicated data collection pipeline. This disaggregated design, with the RAN and 5GC on separate machines, allows for granular analysis of the interfaces between network components.

\subsubsection{User Equipment (UE) Domain}
For our experiments, two UE(s) generates traffic flows for four distinct applications: Live Streaming, Online Gaming, Video Conferencing, and VoIP in 5G SA mode. This controlled generation produces the labeled dataset with known ground-truth classifications required for our supervised ML models.

\subsubsection{Radio Access Network (RAN) Domain}
The RAN domain is architected as a fully software-defined 5G gNodeB (gNB) and serves as the primary data acquisition point for our system. Its design encompasses the complete 3GPP RAN protocol stack, from the PHYsical layer (PHY) to the Radio Resource Control (RRC) layer \cite{srsRAN_project}. The gNB's core responsibility is to manage the radio resources, but crucially for this work, its software-defined nature provides profound visibility into the RAN's internal operations. This allows for the extraction of granular, real-time metrics from all protocol layers, which form the feature set for our ML models.

\subsubsection{5G Core (5GC) Domain}
We incorporated a service-based 5G Core (5GC), adhering to 3GPP specifications, to manage the network's control and user planes. Standard NFs such as the AMF and SMF handle control plane functions like access and session management, while the UPF manages user plane packet forwarding to the external Data Network (DN). The 5GC's primary role in this work is to provide the necessary end-to-end connectivity, creating realistic network conditions for data collection at the RAN.

\subsubsection{Architectural Interfaces and Data Flow}
The RAN and 5GC domains are connected via standard 3GPP interfaces over an IP network. The N2 interface provides the control plane link between the gNB and AMF, while the N3 interface tunnels user plane data between the gNB and UPF using GTP. An end-to-end data flow begins with control signaling on N2 to establish a PDU session, after which user traffic flows across the N3 tunnel to the external DN. This clear separation of functions is fundamental to conducting repeatable experiments on the testbed.

\subsubsection{Data Collection and ML Processing Pipeline}
This component is the cornerstone of our experimental framework. It is designed to systematically capture and process RAN performance data for subsequent ML analysis.

\begin{itemize}
    \item Data Source: The primary source of data is the real-time console trace output from the srsRAN gNB software. This trace contains a rich stream of information from the PHY, MAC, and RLC layers, including metrics such as CQI, MCS, BLER, throughput, and buffer status.
    
    \item Collection Mechanism: A custom Python script runs concurrently with the gNB, parsing the live console output. This script identifies and extracts relevant metric values, associates them with the active traffic class, and formats them into a structured dataset.

    \item ML Model Application: The collected and labeled dataset serves as the input for our machine learning pipeline. Various ML models are trained and tested on this data. The performance of these models is then compared to evaluate their effectiveness in identifying application types based solely on RAN-level metrics.
\end{itemize}

\section{Methodology}

\begin{figure}[htbp]
    \centerline{\includegraphics[width=\linewidth]{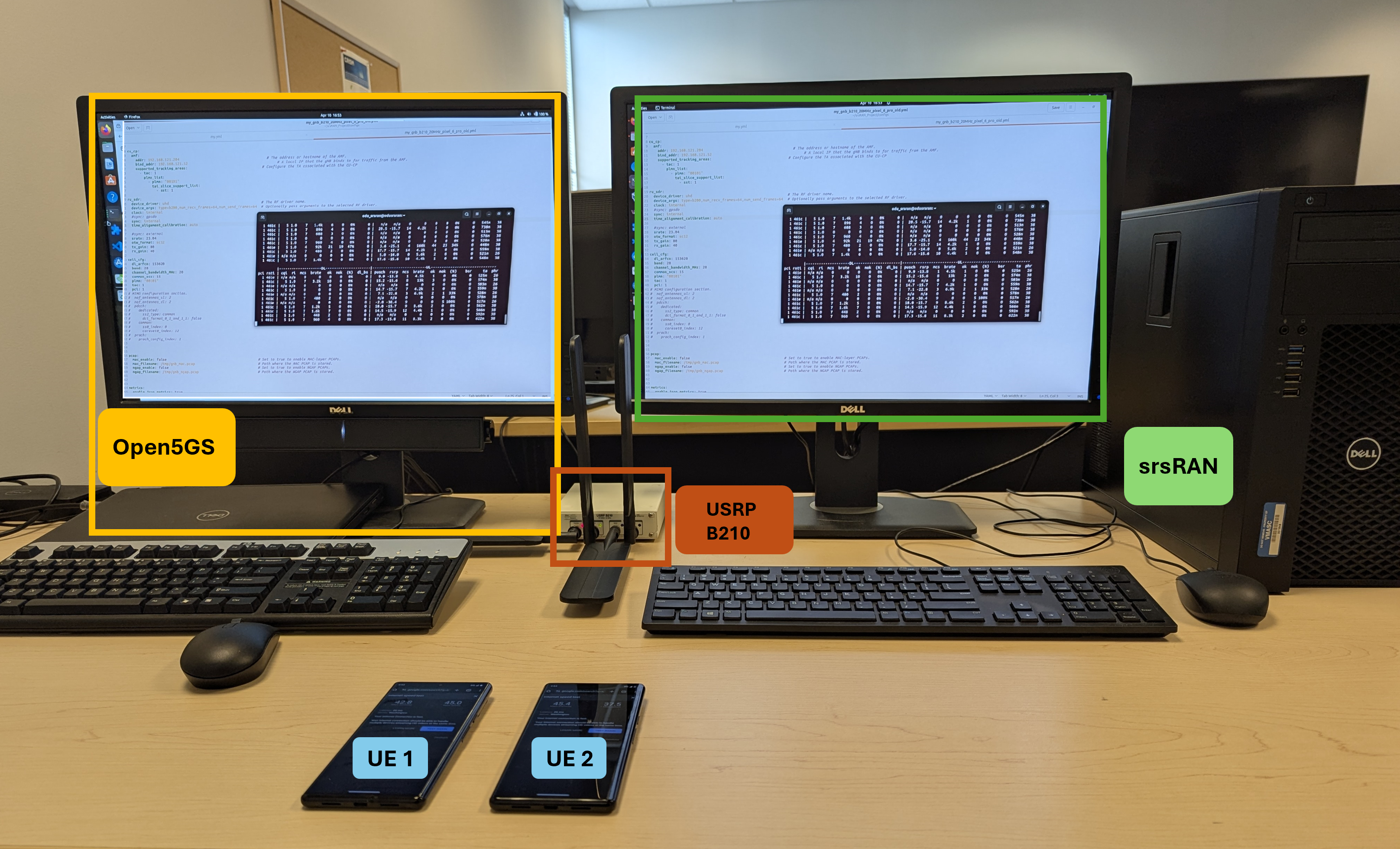}}
    \caption{5G standalone (SA) experimental testbed setup.}
    \label{fig:testbed_label}
\end{figure}
    
\subsection{Opensource-based Experimental Testbed Setup}
To validate our framework, we implemented an end-to-end 5G SA private network in a controlled lab environment. The testbed consists of two interconnected machines running Ubuntu 22.04 LTS: one hosting the 5GC and the other running the gNB, in Fig. 2, ensuring a clear separation of network functions.

Our testbed architecture (Fig. 2) uses open-source software for the 5G Core (5GC), gNB, and User Equipment (UE). The 5GC is implemented with Open5GS (v2.7.5) \cite{open5gs}, an open-source 3GPP-compliant core. We configured all essential Network Functions (NFs), including the AMF, SMF, and UPF. Key configurations involved establishing the control plane connection between the AMF and the gNB, setting up the GTP-U tunnel at the UPF, and using a common PLMN ID (MCC: \texttt{001}, MNC: \texttt{01}) to ensure interoperability with the radio segment.

The Radio Access Network (RAN) was implemented on a dedicated host using the srsRAN Project (release 24.10.1) \cite{srsRAN_project}, with a USRP B210 software-defined radio (SDR) as the RF front-end. The gNB was configured with the 5G Core's AMF address for connectivity. It operated in Frequency Division Duplex (FDD) mode in the sub-6 GHz n28 band with a 20 MHz bandwidth (Downlink: 780.5 MHz, Uplink: 725.5 MHz). The cell was configured with a Tracking Area Code (TAC) of 1 and a Physical Cell ID (PCI) of 1.


\begin{table}[ht]
\centering
\caption{Number of Samples From Two UE(s)}
\label{tab:traffic_class_packets}
\renewcommand{\arraystretch}{1.2} 
\begin{tabular}{l S[table-format=5.0] S[table-format=6.0]}
\toprule
\textbf{Class} & {\textbf{UE-1}} & {\textbf{UE-2}} \\
\midrule
Streaming    & 6334  & 5771 \\
Gaming       & 7798  & 4633 \\
Conference   & 5570  & 5175  \\
VoIP         & 6395  & 5498  \\
\bottomrule
\end{tabular}
\end{table}

Two Google Pixel 6 Pro smartphones were used as UE, each fitted with a programmable SIM card provisioned with network credentials ($IMSI$, $K$, $OP_c$) from the Open5GS database. The UEs successfully attached to the gNB and registered with the 5G Core. End-to-end data plane connectivity was then verified via \texttt{iperf3} tests between a UE and a local server, confirming the testbed's operational stability for experiments.

\subsection{Data Collection Methodology}
To evaluate the network's performance under realistic conditions, we developed a structured methodology for generating diverse user traffic patterns and systematically extracting key performance indicators from the RAN.

\subsubsection{Traffic Generation Scenarios}
To build a robust dataset, we defined four traffic scenarios simulating common user behaviors, each with a distinct network load profile. To ensure data integrity, the UE exclusively generated one traffic type at a time. Each of the four scenarios ran for approximately 3 hours and 20 minutes, with Radio Access Network (RAN) metrics captured at 1 Hz, yielding around 12,000 data samples per category. The dataset is balanced as listed in Table I. 
The four scenarios were:


\begin{enumerate}
    \item[i.] Video Streaming: A 1080p YouTube video was used to generate high, sustained downlink throughput with moderate latency tolerance.

    \item[ii.] Interactive Gaming: Using PUBG Mobile and Roblox, this scenario created bursty, low-latency bidirectional traffic that is highly sensitive to delay and jitter.

    \item[iii.] Video Conferencing: A Microsoft Teams video call produced sustained, symmetric, real-time traffic sensitive to both latency and packet loss.

    \item[iv.] VoIP Call: A Microsoft Teams audio-only call simulated low-bandwidth, periodic traffic that is extremely sensitive to latency and jitter.
\end{enumerate}

\subsubsection{RAN Metrics Extraction}
Real-time performance metrics were collected directly from the srsRAN gNB instance. The srsRAN Project provides a built-in functionality to export a comprehensive set of physical layer (PHY) and Medium Access Control (MAC) layer metrics. We configured the gNB to encapsulate these metrics in JSON format and stream them continuously over a UDP socket to a specified port on the gNB host machine.

Our data collection pipeline was implemented using two custom Python scripts:

\begin{figure}[htbp]
    \centerline{\includegraphics[width=\linewidth]{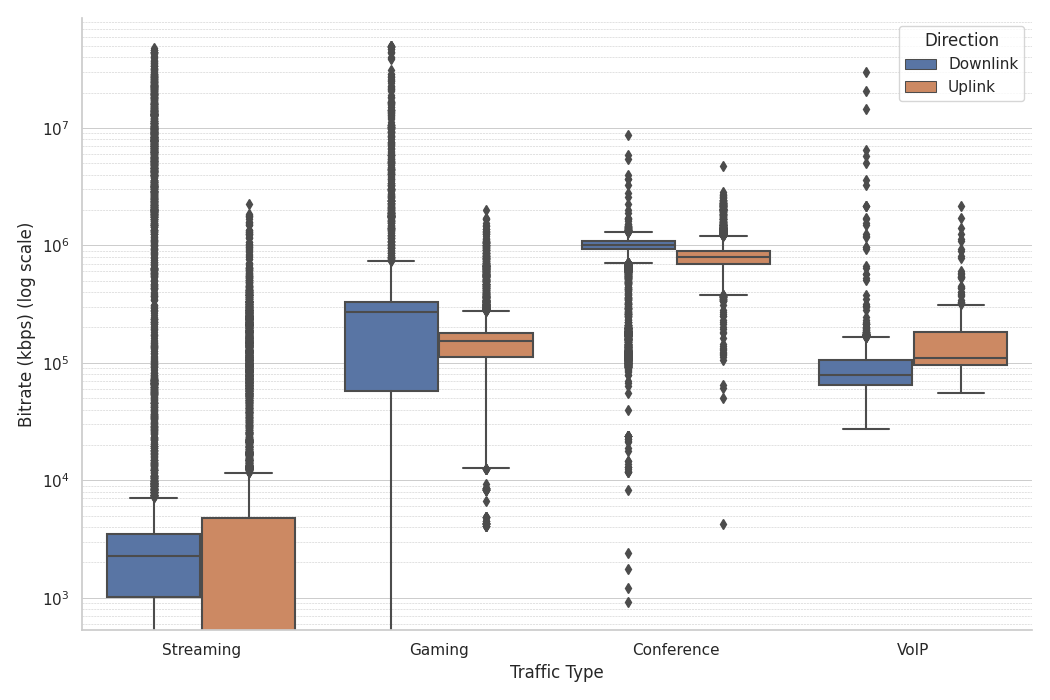}}
    \caption{Bitrate (Downlink/Uplink) data distribution.}
    \label{fig:bitrate_label}
\end{figure}

\begin{itemize}
    \item Data Capture: The first script functioned as a UDP client, persistently listening on the designated port for incoming data from the srsRAN gNB. Its sole responsibility was to receive the JSON-formatted metric reports and write them sequentially into a log file, creating a raw, time-ordered record of the RAN's state during each traffic scenario.

    \item Data Processing: After the completion of an experimental run, a second Python script was used to parse the generated JSON log file. This script extracted key performance indicators of the gNB. It then flattened the nested JSON structure and converted the time-series data into a tabular comma-separated values (CSV) format.
\end{itemize}

\subsubsection{Dataset Description}
We collect a comprehensive set of time-series data from the 5G gNodeB, that captures the state of the physical (PHY) and Medium Access Control (MAC) layers. These metrics serve as input features for our machine learning model and are discussed as follows. 

We selected a total of thirteen (13) features that are listed in Table II. These metrics were chosen to capture the unique data signatures and QoS profiles inherent to each application. We utilize downlink and uplink metrics, including bitrate, Modulation and Coding Scheme, and transport block success/failure counts, to characterize throughput demands and link reliability. These features are essential for distinguishing high-bandwidth, asymmetric traffic like streaming from the symmetric, interactive patterns of conferencing and gaming. Crucially, cell latency is included as a primary QoS differentiator to identify delay-sensitive applications like gaming and VoIP.

To complement the data flow metrics, we incorporate features that reflect the underlying radio link conditions and UE transmission intent. The PUSCH and PUCCH Signal-to-Noise Ratios provide a direct measure of channel quality, which governs overall link performance. Furthermore, the Buffer Status Report offers an explicit indication of the volume of data pending in the UE's uplink buffer, serving as a powerful signal for identifying applications that actively generate data. Collectively, this feature set provides a robust signature of RAN behavior, enabling our ML models to accurately discern the application type based on its network footprint.


\begin{table}[!t]
\centering
\caption{Key RAN Features for our Experiment}
\label{tab:network_features}
\renewcommand{\arraystretch}{1.2} 
\begin{tabularx}{\columnwidth}{l c c l X}
\toprule
\textbf{Layer} & \textbf{DL} & \textbf{UL} & \textbf{Feature} & \textbf{Description} \\
\midrule
\multirow{3}{*}{PHY}
 & \checkmark & \checkmark & mcs & Modulation and Coding Scheme \\

\cmidrule(lr){2-5}
 &            & \checkmark & pucch sinr & SINR on the Physical Uplink Control Channel \\
 
\cmidrule(lr){2-5} 
 &            & \checkmark & pusch sinr & SINR on the Physical Uplink Shared Channel \\

\midrule 
\multirow{5}{*}{MAC}
 & \checkmark &            & bs & Downlink transport block size in bits \\
\cmidrule(lr){2-5}
 &            & \checkmark & bsr & UE buffer status report for uplink data \\
\cmidrule(lr){2-5}
 & \checkmark & \checkmark & latency & Round-trip time within the cell \\
\cmidrule(lr){2-5}
 & \checkmark & \checkmark & ok pkt & Successfully sent packets \\
\cmidrule(lr){2-5}
 & \checkmark & \checkmark & nok pkt & Dropped packets \\
\cmidrule(lr){2-5}
  & \checkmark & \checkmark & brate & Bitrate (bits/sec) \\ 
\bottomrule
\end{tabularx}
\end{table}

\subsection{Data Preprocessing and Preparation}
A preliminary exploratory data analysis revealed that several metrics (\texttt{PCI, CQI, RNTI, RI}) remained constant throughout all experimental scenarios. Fig. 3 describes distribution of DL/UL data where we can see separation of different category. Metrics that exhibited zero variance across the entire data collection campaign were systematically excluded, as constant features provide no discriminatory power for machine learning classification. Consequently, they were removed from the final feature set to reduce model complexity and avoid unnecessary computational overhead without any loss of predictive capability. Timing Advance metrics were excluded because they represent the UE's physical distance from the gNB for synchronization, which is uncorrelated with the application's traffic profile. Since a user can run any application at any distance, these metrics have no discriminatory power for classifying traffic types.


Our dataset comprises 47,175 instances, each defined by 13 RAN performance features. The data is well-balanced across four distinct application categories: Streaming, Gaming, Conference, and VoIP, with each class representing approximately 25\% of the total samples. An initial inspection confirmed the dataset's integrity, revealing no missing values or outliers, thus bypassing the need for data imputation or extensive cleaning.

The raw dataset extracted from the srsRAN next-generation NodeB (gNB) consists of dynamic, high-resolution time-series metrics reflecting network state and user equipment (UE) interactions. To ensure the machine learning models learn generalizable network behaviors rather than memorizing highly correlated adjacent data points, a strict preprocessing and partitioning pipeline was implemented.

\subsubsection{Data Cleaning and Normalization}
Initial preprocessing involved the removal of anomalous spikes caused by logging errors and the imputation of missing values using forward-filling to maintain the temporal continuity of the RAN metrics. Given the heterogeneous nature of the extracted features—ranging from low-variance signal-to-noise ratios (SNR) to high-variance byte counts—feature scaling was strictly applied. All features were normalized to a standard scale using Z-score standardization, ensuring that gradient-based models, such as the Multi-Layer Perceptron (MLP) and Long Short-Term Memory (LSTM) networks, converged efficiently without being biased by features with larger magnitudes.
\subsubsection{Eradication of Temporal Data Leakage}
In preliminary evaluations, standard randomized stratified splitting yielded artificially inflated classification validation accuracies exceeding $99\%$. Because RAN telemetry data is intrinsically sequential, standard random sampling induces severe data leakage; adjacent, highly autocorrelated time-steps bleed across the training and testing sets, allowing models to interpolate rather than extrapolate.To absolutely eradicate this leakage, we discarded randomized splitting in favor of a strict temporal and session-based partitioning strategy. The dataset was divided into contiguous, non-overlapping temporal blocks representing entirely distinct network sessions.
\subsubsection{Construction of Zero-Shot and Few-Shot Evaluation Sets}
Based on the session-based partitioning, the data was structured to support two distinct evaluation paradigms:

\begin{itemize}
    \item Zero-Shot Baseline (Source Distribution): The training and validation sets were constructed exclusively from historical network sessions. The test set was formulated from entirely unseen, future network sessions. This guarantees a true `Zero-Shot' evaluation environment where the test accuracy precisely reflects the model's pure generalization capabilities to novel network states.
    
    \item Few-Shot Calibration Set (Target Distribution): To support the Few-Shot calibration phase, a highly restricted, randomly sampled subset of $k$ instances (where $k$ represents the `shots') was extracted from the initial timeframe of the unseen test sessions. This isolated calibration set allows the models to perform domain adaptation without violating the integrity of the broader unseen test set.
\end{itemize}

\subsection{Machine Learning Models}
To comprehensively evaluate the robustness of anomaly detection and classification within the srsRAN environment, we deployed a diverse portfolio of algorithms. These range from highly interpretable classical machine learning models to complex deep learning architectures capable of capturing non-linear and sequential dependencies. 

\subsubsection{Classical Machine Learning Algorithms}
We established a rigorous baseline using several traditional algorithms known for their varying trade-offs between computational efficiency and predictive power:
\begin{itemize}
    \item Tree-Based Ensembles: XGBoost and Random Forest were deployed due to their robust performance on tabular data and resilience to feature scaling variances. XGBoost, in particular, leverages gradient boosting to sequentially minimize classification errors. A standard Decision Tree was also included as a high-interpretability, low-latency baseline.

    \item Distance and Margin-Based Classifiers: Support Vector Machines (SVM) utilizing an RBF kernel, K-Nearest Neighbors (KNN), and Logistic Regression were evaluated to assess how well the feature space could be linearly or non-linearly separated using traditional mathematical boundaries.
\end{itemize}

\subsubsection{Deep Learning Architectures}
To address the highly dimensional and temporally correlated nature of the RAN telemetry, two distinct neural network architectures were implemented:

\begin{itemize}
    \item Multi-Layer Perceptron (MLP): A fully connected feed-forward neural network was utilized to capture complex, non-linear feature interactions that classical boundary models might miss. The MLP serves as our primary deep learning baseline and proved highly responsive to our few-shot calibration techniques.
    \item Long Short-Term Memory (LSTM) Networks: Recognizing that the srsRAN data consists of contiguous time-series metrics, an LSTM architecture was introduced.
\end{itemize}

Unlike standard feed-forward networks, LSTMs utilize a proprietary cell state regulated by input, output, and forget gates. This allows the network to maintain memory of long-term temporal dependencies within the network sessions, making it theoretically well-suited for predicting sequential RAN behaviors without suffering from the vanishing gradient problem.

\subsection{Experimental Setup and Evaluation Protocol}

To ensure reproducibility and a fair comparison across all architectures, a standardized training and evaluation pipeline was established. All computational experiments and model training were conducted on the High-Performance Computing (HPC) cluster provided by Old Dominion University. The allocated experimental environment utilized a dedicated compute node equipped with an NVIDIA V100 GPU, 8 CPU cores, and 144 GB of RAM. Algorithm implementation and hardware-accelerated processing were executed using Python 3.10.14 and the TensorFlow framework (version 2.16.2).

\subsubsection{Hyperparameter Optimization}
Rigorous hyperparameter tuning was conducted for each model to ensure optimal baseline performance. We utilized GridSearchCV with 3-fold cross-validation exclusively on the source training distribution. This strict isolation ensures that the validation process did not inadvertently expose the models to the target test distribution, maintaining the integrity of our leakage-free methodology. Model selection during this phase was based on the highest classification accuracy achieved on the held-out validation folds.

\subsubsection{Phase 1: Zero-Shot Evaluation Protocol}
To establish a rigorous baseline for model generalization, we first executed a Zero-Shot evaluation. In this paradigm, models were trained and tuned entirely on historical network sessions (the source domain). They were then evaluated on completely distinct, future network sessions (the target domain) without any prior exposure or structural adaptation.

\begin{table}[htbp]
\centering
\caption{Zero-Shot Baseline Performance Metrics.}
\label{tab:model_performance_single_column}
\renewcommand{\arraystretch}{1.5} 

\resizebox{\columnwidth}{!}{%
\begin{tabular}{@{} l S[table-format=2.2] S[table-format=2.2] S[table-format=1.4] S[table-format=2.2] S[table-format=2.2] @{}}
\toprule
\textbf{Model Name} & {\makecell{Tune Time\\(s)}} & {\makecell{Train Time\\(s)}} & {\makecell{Inference\\(ms/Sample)}} & {\makecell{Validation\\Acc. (\%)}} & {\makecell{Test Acc.\\(\%)}} \\
\midrule
XGBoost & 5.60 & 0.57 & 0.0017 & 99.50 & 72.87 \\
Random Forest & 11.97 & 6.16 & 0.0082 & 99.46 & 69.17 \\
Decision Tree & 0.80 & 0.16 & 0.0001 & 99.02 & 65.36 \\
K-Nearest Neighbors & 2.96  & 0.05 & 0.1941 & 98.03 & 64.01 \\
MLP Classifier & 38.08 & 8.96 & 0.0010 & 99.27 & 59.72 \\
LSTM & {NA} & 94.97 & 0.1055 & 99.18 & 59.08 \\
Logistic Regression & 27.47 & 13.09 & 0.0005 & 95.10 & 57.48 \\
SVM (RBF Kernel) & 4.66 & 0.84 & 0.0519 & 98.95 & 57.24 \\
\bottomrule
\end{tabular}%
}
\end{table}

This protocol tests the `pure' generalization capability of the algorithms. Because RAN environments are highly dynamic, a model trained on one temporal state may suffer a performance drop when exposed to a drifted network state. The Zero-Shot baseline empirically quantifies this generalization gap, providing a realistic measure of how these models would perform if deployed in a production gNB without continuous retraining.

\subsubsection{Phase 2: Few-Shot Calibration Protocol}
To bridge the generalization gap observed in the Zero-Shot phase, we introduced a Few-Shot calibration protocol. This phase simulates a rapid, lightweight domain adaptation process suitable for dynamic RAN environments.

A highly constrained subset of $k$ samples (the `shots') was extracted from the initial timeframe of the target test sessions. For classical machine learning algorithms, the models were retrained using the augmented dataset comprising the original training data and the $k$-shot calibration samples. For the deep learning architectures (MLP and LSTM), the pre-trained Zero-Shot models were fine-tuned for a limited number of epochs using only the $k$-shot target samples, allowing the networks to rapidly adjust their decision boundaries to the new network state. The calibrated models were then evaluated on the remainder of the unseen test set, establishing the effectiveness of Few-Shot learning in recovering classification accuracy in shifted RAN environments.



\begin{table}[htbp]
\centering
\caption{Few-Shot Calibration Performance Metrics.}
\label{tab:few_shot_performance_single_column}
\renewcommand{\arraystretch}{1.5} 

\resizebox{\columnwidth}{!}{%
\begin{tabular}{@{} l S[table-format=2.2] S[table-format=2.2] S[table-format=1.4] S[table-format=2.2] S[table-format=2.2] @{}}
\toprule
\textbf{Model Name} & {\makecell{Tune Time\\(s)}} & {\makecell{Train Time\\(s)}} & {\makecell{Inference\\(ms/Sample)}} & {\makecell{Validation\\Acc. (\%)}} & {\makecell{Test Acc.\\(\%)}} \\
\midrule
MLP Classifier & 38.74 & 7.58 & 0.0006 & 99.27 & 86.25 \\
XGBoost & 6.07 & 0.61 & 0.0017 & 99.58 & 82.60 \\
Random Forest & 13.76 & 7.18 & 0.0095 & 99.43 & 79.90 \\
LSTM & NA & 72.23 & 0.1168 & 99.33 & 77.59 \\
K-Nearest Neighbors & 3.19  & 0.05 & 0.2063 & 97.87 & 76.32 \\
Decision Tree & 0.89 & 0.19 & 0.0001 & 98.83 & 74.67 \\
SVM (RBF Kernel) & 5.42 & 1.06 & 0.0625 & 99.00 & 73.39 \\
Logistic Regression & 15.81 & 0.39 & 0.0005 & 96.40 & 72.25 \\
\bottomrule
\end{tabular}%
}
\end{table}

\section{Experimental Results and Discussion}
To comprehensively evaluate the robustness of the proposed models within the srsRAN environment, we analyzed their performance across two distinct experimental paradigms: a strictly isolated Zero-Shot baseline and a targeted Few-Shot calibration. Performance was measured not only by validation and test accuracy, but also through computational metrics critical for real-time Radio Access Network (RAN) deployment, specifically tuning time, training time, and inference latency.

\subsection{Zero-Shot Generalization Performance}

To establish a rigorous baseline and ensure the complete elimination of temporal data leakage, our first experimental phase evaluated all models under a strict zero-shot protocol. In this paradigm, models were trained and hyperparameter-tuned exclusively on the source distribution and subsequently evaluated on a strictly segregated, unseen test distribution derived from distinct srsRAN sessions. This approach prevents highly correlated adjacent time-series samples from artificially inflating model performance. The comprehensive baseline metrics are presented in Table III.

The results reveal a significant generalization gap across all evaluated architectures. While validation accuracies consistently approached or exceeded 99\%, the test accuracies degraded substantially when exposed to the out-of-distribution RAN metrics. This stark contrast highlights the inherent volatility of dynamic O-RAN environments and validates our rigorous session-based data splitting methodology.

In this purely zero-shot setting, tree-based ensemble methods demonstrated superior robustness compared to neural architectures. XGBoost emerged as the optimal classifier, achieving the highest test accuracy of 72.87\%, followed closely by Random Forest at 69.17\%. The decision boundary constructions of these ensemble methods appear more resilient to the covariate shifts present in the unseen test data.

Conversely, the neural network models—specifically the MLP Classifier (59.72\%) and the LSTM (59.08\%)-suffered severe performance degradation. Despite their capacity to map complex, non-linear relationships during training (evidenced by their $>$ 99\% validation scores), their highly parameterized nature made them overly sensitive to the specific statistical properties of the training distribution, leading to poor zero-shot generalization.

However, it is critical to note that the inference latency for the top-performing XGBoost model remained exceptionally low at 0.0017 ms/sample, confirming its viability for real-time MAC-layer deployment despite the drop in raw accuracy. This observed generalization gap directly motivates the necessity of our subsequent few-shot calibration phase, which aims to recover the performance of these models in target environments.

\subsection{Few-Shot Calibration Performance}
To address the significant performance degradation observed in the zero-shot evaluation, we introduced a few-shot calibration protocol. In highly dynamic srsRAN environments, relying solely on source-domain training is often insufficient due to inevitable distribution shifts in target network conditions. By exposing the models to a highly constrained subset ($k$-shots) of labeled data from the target domain, we aimed to calibrate their decision boundaries to the new statistical distribution without requiring complete retraining. The performance metrics following this calibration phase are detailed in Table IV.

    
        

\begin{figure*}[htbp]
    \centering
    
    \subfloat[XGBoost Model result.]{\label{fig:xgboost_dist}%
        \includegraphics[width=0.45\textwidth]{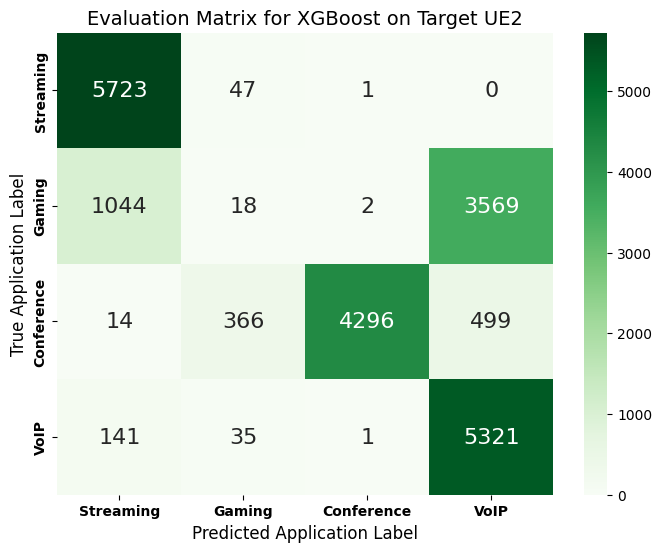}}
    \hfill
    \subfloat[MLP Model result.]{\label{fig:MLP_dst}%
        \includegraphics[width=0.45\textwidth]{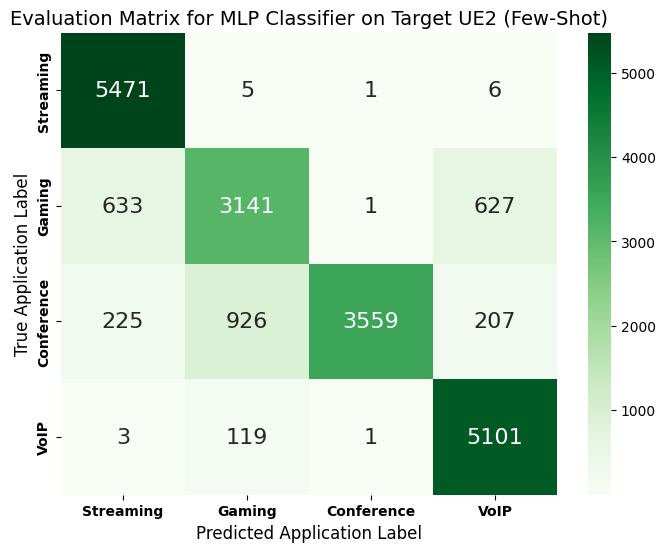}}
        
    \caption{Confusion Matrices for the XGBoost and MLP models.}
    \label{fig:confusion_comparison}
\end{figure*}

The application of few-shot calibration yielded a profound restructuring of the performance hierarchy, most notably among the neural network architectures. The MLP Classifier exhibited an extraordinary recovery, from a zero-shot test accuracy of 59.72\% to a leading 86.25\%. Similarly, the LSTM network improved its test accuracy from 59.08\% to 77.59\%.

This dramatic leap demonstrates that neural architectures, while highly susceptible to covariate shifts in pure zero-shot scenarios, possess a superior capacity for rapid domain adaptation. The few-shot samples provided sufficient gradient signals to meaningfully adjust the network weights, allowing the MLP to realign its complex, non-linear representations with the target srsRAN distribution.

Classical ensemble methods also benefited from calibration, though to a lesser mathematical degree. XGBoost improved from 72.87\% to 82.60\%, and Random Forest increased from 69.17\% to 79.90\%. While these tree-based models established a higher initial zero-shot baseline, their rigid node-splitting structures inherently limit the extent to which a small handful of new samples can reshape the overall decision space compared to the fluid weight adjustments in an MLP.

Crucially, the MLP Classifier achieved this superior accuracy while maintaining an exceptionally low inference latency of 0.0006 ms/sample. This makes the calibrated MLP an optimal candidate for deployment within tight O-RAN Real-Time Intelligent Controller (RIC) control loops, where sub-millisecond execution is mandatory. In contrast, while the LSTM achieved respectable accuracy, its higher inference latency of 0.1168 ms/sample and extended training time (72.23 s) suggest it may be better suited for non-real-time analytical tasks rather than immediate MAC-layer scheduling.

\subsection{Computational Complexity and Latency Trade-offs}

In the context of Open Radio Access Network (O-RAN) architectures, achieving high classification accuracy is only one facet of model viability. Algorithms deployed within the Real-Time Intelligent Controller (RT-RIC) or directly at the MAC layer must adhere to ultra-low latency constraints, often requiring sub-millisecond execution times. Therefore, evaluating the trade-offs between model accuracy, training overhead, and inference latency is critical for practical deployment.

An analysis of the inference latencies (recorded in milliseconds per sample) reveals distinct operational tiers among the evaluated models. The Decision Tree achieved the lowest absolute inference time at 0.0001 ms/sample in both protocols, functioning essentially as a series of simple threshold evaluations. However, its calibrated accuracy plateaued at 74.67\%, making it less desirable when higher reliability is required.

The Few-Shot MLP Classifier presents the most compelling operational profile. Despite achieving the highest calibrated test accuracy (86.25\%), it maintained an exceptionally low inference latency of 0.0006 ms/sample. Once the network weights are optimized during the few-shot phase, forward propagation through the perceptron layers involves highly parallelizable matrix multiplications, rendering it highly efficient for real-time RAN scheduling. Similarly, the zero-shot baseline champion, XGBoost, maintained a highly viable inference latency of 0.0017 ms/sample, positioning it as an excellent, deployment-ready fallback when target-domain calibration data is unavailable.

Conversely, models reliant on sequential processing or extensive memory lookups demonstrated prohibitive latencies for strict real-time applications. The K-Nearest Neighbors (KNN) algorithm exhibited the highest inference latency at 0.2063 ms/sample (few-shot), as it requires computing distances against the stored training manifold during every prediction. The LSTM network, while demonstrating strong accuracy recovery (77.59\%), required 0.1168 ms/sample for inference. The recurrent nature of the LSTM, which prevents parallelization of sequential time-steps, introduces a processing bottleneck. Furthermore, the LSTM imposed the highest computational burden during the training phase, consuming 72.23 s for few-shot calibration compared to just 7.58 s for the MLP.

Ultimately, these results indicate that while sequential models like LSTMs may be suitable for Non-Real-Time (Non-RT) RIC analytics or long-term network monitoring, feedforward architectures like the calibrated MLP or tree-based ensembles like XGBoost are the optimal architectures for latency-critical, real-time srsRAN control loops.

\subsection{Class-Level Generalization and Decision Boundary Analysis}

To understand the underlying mechanics of the generalization gap and the subsequent few-shot recovery, we conducted a class-level performance analysis using confusion matrices for our baseline champion (XGBoost) and our calibrated champion (MLP Classifier) on the Target UE2 data. The results, illustrated in Figure 4, reveal significant disparities in how these models process specific network traffic profiles under distribution shifts.

\begin{enumerate}
    \item \textbf{Zero-Shot XGBoost Class Imbalance:} Figure 4(a) illustrates the evaluation matrix for the zero-shot XGBoost model. While it achieved an overall accuracy of 72.87\%, the matrix exposes a severe class-level failure: the model was entirely unable to generalize the "Gaming" traffic profile in the target environment. Out of all actual Gaming samples, XGBoost correctly classified only 18 instances. It overwhelmingly misclassified Gaming traffic as VoIP (3,569 instances) and Streaming (1,044 instances).

    This indicates that in a pure zero-shot scenario, the statistical signatures of Gaming traffic (e.g., packet inter-arrival times, payload sizes) on Target UE2 shifted so dramatically from the source training data that they became indistinguishable from VoIP and Streaming within XGBoost's rigid decision tree boundaries. The model achieved its baseline accuracy primarily by accurately predicting the remaining three classes (Streaming, Conference, and VoIP), masking a critical failure in minority-class recall.
    
    \item \textbf{Few-Shot MLP Calibration and Recovery:} Figure 4(b) demonstrates the profound impact of the few-shot calibration protocol on the MLP Classifier. By introducing a minimal $k$-shot sample of the target distribution during training, the neural network was able to dynamically adjust its weights to learn the new, shifted signature of the Gaming class.

    The recovery is substantial: the MLP correctly identified 3,141 Gaming instances, compared to XGBoost's 18. While this recalibration introduced minor confusion between Conference and Gaming traffic (926 Conference instances misclassified as Gaming), the overall macro-level performance became significantly more balanced. The MLP maintained high true-positive rates for Streaming (5,471), Conference (3,559), and VoIP (5,101) while successfully rescuing the Gaming class from complete misclassification.

    Ultimately, this analysis proves that while classical ensemble methods like XGBoost may provide a higher gross accuracy in uncalibrated zero-shot scenarios, their rigid boundaries are prone to catastrophic failures on shifted individual classes. The Few-Shot MLP not only provides superior overall accuracy (86.25\%) but guarantees a far more reliable, balanced detection capability across all critical 5G application types.
\end{enumerate}

\section{Conclusion And Future Work}

This paper presents a privacy-preserving, machine learning-driven approach to 5G traffic classification using exclusively RAN metrics. Utilizing a custom 5G Standalone (SA) testbed powered by srsRAN and Open5GS , we generated a novel dataset  and addressed a pervasive methodological flaw in RAN traffic analysis: temporal data leakage that artificially inflates model performance.

By enforcing a strict zero-shot evaluation protocol, we quantified the true generalization gap in dynamic networks. Our baseline analysis showed that while classical tree-based models like XGBoost achieved a moderate baseline test accuracy of 72.87\%, their rigid decision boundaries caused severe class-level failures, completely failing to identify Interactive Gaming traffic under distribution shifts.

To resolve this, we introduced a few-shot calibration framework for rapid domain adaptation. By leveraging merely a 5\% subset of target-domain data, this approach successfully recalibrated the Multi-Layer Perceptron (MLP) architecture, elevating its test accuracy to a leading 86.25\% and fully recovering minority-class detection capabilities. Crucially, the calibrated MLP achieves this balanced accuracy while maintaining an ultra-low inference latency of 0.0006 ms/sample, validating its operational feasibility for real-time deployment in O-RAN Intelligent Controllers (RIC).

Future work will focus on integrating this calibrated MLP as a functioning xApp within an O-RAN architecture, expanding the dataset for highly multiplexed multi-UE scenarios, and extending this few-shot methodology toward RAN-level anomaly detection.

\bibliographystyle{IEEEtran}
\bibliography{references.bib}

\end{document}